\documentclass[aps,groupedaddress,preprint,superscriptaddress]{revtex4-1}

\usepackage{amsfonts}
\usepackage{graphicx}
\usepackage{mathcomp}
\usepackage{wasysym}
\begin{document}
\title{Etch Induced Microwave Losses in Titanium Nitride Superconducting Resonators }\homepage{ This contribution of NIST, an agency of the U.S. government, is not subject to copyright}

\author{Martin Sandberg}

\email[]{Electronic addreass: martin.sandberg@nist.gov}

\author{Michael R. Vissers}


\author{Jeffrey S. Kline}


\author{Martin Weides}

\altaffiliation{Current address: Karlsruhe Institute of Technology, 76131 Karlsruhe, Germany}

\author{Jiansong Gao}

\affiliation{National Institute of Standards and Technology, Boulder, Colorado 80305, USA}

\author{David S. Wisbey}

\affiliation{Saint Louis University, St. Louis, Missouri 63103, USA}

\author{David P. Pappas}

\email[]{Electronic address: david.pappas@nist.gov}
\affiliation{National Institute of Standards and Technology, Boulder, Colorado 80305, USA}

\begin{abstract}
We have investigated the correlation between the microwave loss and patterning method for coplanar waveguide titanium nitride resonators fabricated on Si wafers. Three different methods were investigated: fluorine- and chlorine-based reactive ion etches and an argon-ion mill. At high microwave probe powers the reactive etched resonators showed low internal loss, whereas the ion-milled samples showed dramatically higher loss. At single-photon powers we found that the fluorine-etched resonators exhibited substantially lower loss than the chlorine-etched ones. We interpret the results by use of numerically calculated filling factors and find that the silicon surface exhibits a higher loss when chlorine-etched than when fluorine-etched. We also find from microscopy that re-deposition of silicon onto the photoresist and side walls is the probable cause for the high loss observed for the ion-milled resonators.   
      
\end{abstract}

\maketitle

Superconducting resonators are essential building blocks of quantum electrical circuits. They are used for dispersive readout and coupling of superconducting quantum bits, quantum information storage and for detector detector applications \cite{Wallraff2004,Leek2010,DiCarlo2010,Mariantoni2011,Day2003}. This has lead to extensive efforts to understand and minimizing the loss of these devices\cite{Barends2010,Sage2011,Wang2009,Khalil2011,Gao2008a,Wenner2011}. One very promising material for building low-loss superconducting resonators is titanium nitride (TiN), which has been shown to have very low loss at high as well as at low drive powers\cite{Leduc2010,Vissers2010}.

Here we present experimental results relating the observed microwave loss in thin-film TiN  resonators to the method of etching used for patterning the devices. The resonators used were frequency-multiplexed quarter-wavelength coplanar waveguides (CPW) on a silicon substrate. On each chip, ten resonators with varying coupling strength and resonance frequencies were coupled to a common feedline\cite{Pappas2011}. The coupling was designed to give an external quality factor ranging from 0.5 million to 5 million, and a resonance frequency between 4 GHz and 7 GHz, depending on kinetic inductance. 

In this work, three different etches were investigated: a fluorine (F)-based reactive ion etch (RIE), a chlorine (Cl)-based RIE and  an argon ion mill. The etches were chosen due to their wide use and fundamentally different natures. The F etch has a relatively low etch rate of TiN compared to its Si etch rate, the Cl etch has almost identical etch rates for Si and TiN, and the argon-ion mill is a completely physical etch.
  
All devices were fabricated on highly resistive intrinsic Si(100) 3" wafers. The wafers were exposed to a hydrofluoric (HF) vapor etch to remove the native oxide prior to the film growth. The HF etch also hydrogen-terminates the Si surface, which has been shown to be crucial for achieving low loss in resonators on Si\cite{Wisbey2010}. Within minutes after the HF etch, the wafers were transferred into a high-vacuum sputtering system. A TiN film (40 nm thick) was deposited by reactive sputtering at 500 $\tccentigrade$, 250 W DC power, 4 mT chamber pressure, 15 sccm argon flow, 10 sccm nitrogen flow, and an RF-induced 100 V substrate DC bias. The TiN films were patterned into CPW resonators by the use of optical lithography and etched by the use of one of the three different methods. Parameters for the etches are summarized in Table \ref{EtchTab}. 

The wafers were diced into chips and wire-bonded into an aluminium sample box equipped with microwave launchers. To reduce the risk of trapping magnetic flux during cool-down, the sample box was placed inside niobium and cryoperm shields. The devices were then cooled to the base temperature of an adiabatic demagnetization refrigerator ($\approx$ 50 mK). The microwave line used to drive the resonators was attenuated by 80 dB and filtered by the use of low-pass filters with a cut-off frequency of 12 GHz. A high-electron-mobility transistor (HEMT) placed at the 3 K stage was used to amplify the signal from the sample. To reduce thermal radiation from the HEMT, a 3.5 GHz high- and a 12 GHz low-pass filter as well as a two stage isolator with an isolation of $\approx$ 40 dB were placed between the sample and the HEMT.           
The microwave response of each device was measured through the scattering parameter $S_{21}$ of the feedline by the use of a vector network analyzer. The internal and external quality factors ($Q_{int}$ and $Q_c$ respectively) were extracted through a circular fitting procedure \cite{Gao2008b} of the real and imaginary parts of the $S_{21}$ response. To maximize the fit accuracy, resonators with similar $Q_{int}$ and $Q_{c}$ were chosen. 

\begin{figure}[h]
\includegraphics[width=8cm]{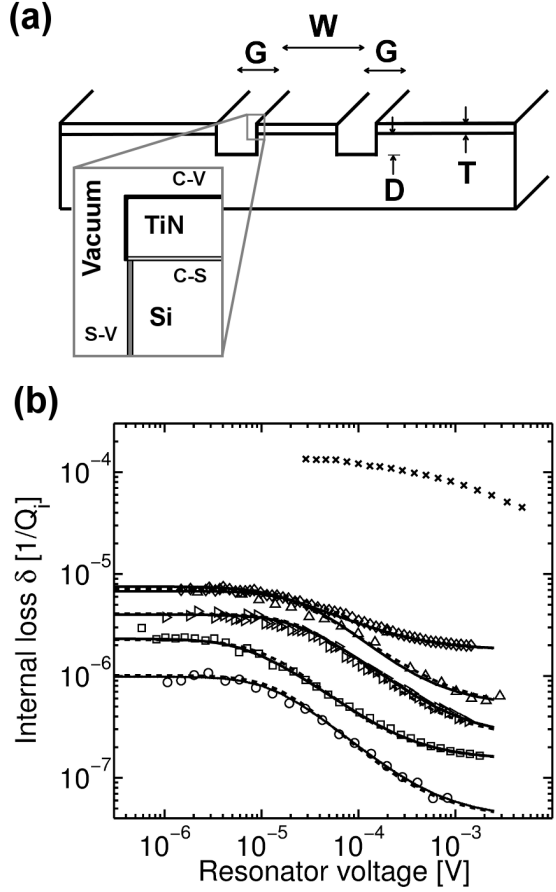}
\caption{\label{lossFig} (a) Sketch of a coplanar waveguide structure. Here G denotes the gap between the ground plane and the centerstrip, W the width of the center strip, D the depth of the trench, and T the thickness of the TiN film. The inset shows the top corner of the CPW center strip and illustrates the position of substrate-vacuum (S-V), conductor-vacuum (C-V), and conductor-substrate (C-S) interfaces in the cross-section of the CPW. (b) Extracted internal loss as a function of internal voltage of the resonators described in Table \ref{ResTab}. The different markers represent the different resonators: ($\ocircle$) Fl1, ($\Box$) Fl2,($\rhd$) Cl1, ($\triangle$) Cl2, ($\Diamond$) Cl3 and ($\times$) IM. The lines are fits of Eqn.\ref{eqLoss} by the use of the calculated electric field in region S-V (solid) and region C-S (dashed).}
\end{figure}

Measurements of six resonators, labelled F1, and F2 for the F-etched, Cl1, Cl2, and Cl3 for the Cl-etched, and IM for the ion-milled, respectively, are presented here. The parameters of the resonators are summarized in Table \ref{ResTab}. The extracted internal loss ($\tan(\delta_{int})=1/ Q_{int}$) is plotted in Figure \ref{lossFig} as a function of the internal resonator voltage. It is clear that the ion-milled resonator, IM, shows substantially higher internal loss compared to the RIE etched resonators. Among the RIE treated resonators it is also clear that F etched resonators have the lowest internal loss. 

The power dependence of the internal loss results from two level systems (TLSs). There are three different surfaces where the TLSs are likely to be located \cite{Gao2008a}: at the substrate-vacuum (S-V) interface in the CPW gap, the conductor-vacuum (C-V) interface, and the conductor-substrate (C-S) interface. The different surfaces are depicted in the inset of Fig. \ref{lossFig} (a). The loss contribution $\delta_V$, from a volume $V$ with dielectric constant $\epsilon_V$, is obtained as\cite{Khalil2011}:
\begin{equation}
\label{eqLoss}
\tan(\delta_{V})=
 \tanh(\frac{\hbar\omega_r}{2k_BT})\frac{\tan(\delta_{V}^{0})\int_{V}\epsilon_{V} |E(\vec{r})|^2\left(1+\left(\frac{E(\vec{r})}{E_c}\right)^2\right)^{-1/2} \mathrm{d}v}{\int_{V_{tot}}\epsilon(\vec{r})|E(\vec{r})|^2 \mathrm{d}v} 
\end{equation}
where $\vec{r}$ is the position inside the volume under integration and $V_{tot}$ is the total volume. Here $E_c$ is the saturation electrical field of the TLSs and $E(\vec{r})$ is the local electric field strength, $\delta_V^0$ is the loss at small electric field and low temperature, and $\omega_r=2\pi f_r$ is the (angular) resonance frequency. The device temperature $T$ was found to be low enough that thermal excitation of the TLSs was negligible, \it i.e., \rm $T<\hbar\omega_r/2k_B$.  

\begin{figure}[h]
\includegraphics[width=8cm]{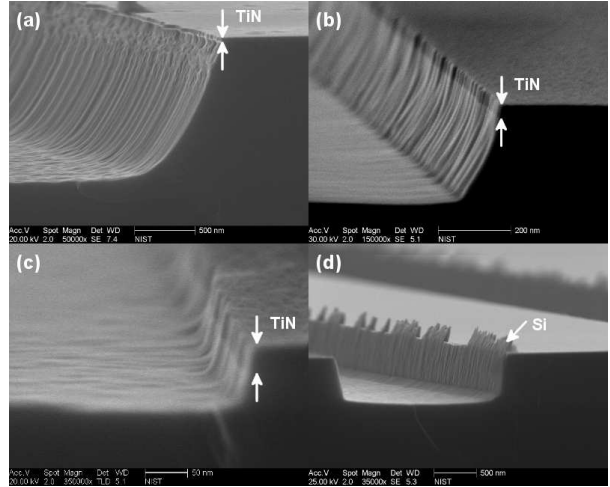}
\caption{\label{SEMfig}SEM images of different etched samples. (a) Trenched F-etch Fl1. (b) Trenched Cl-etch Cl1. (c) Non trenched Cl-etch Cl3. (d) Trenched ion milled IM. From the cross sections it is clear that the F-etched resonator has an undercut profile that is not observed for the Cl-etched resonators. It can also be seen that the ion-milled profile has re-deposited material on the top and sides of the TiN film.}
\end{figure}

To fit Eqn. \ref{eqLoss} to the measured loss, the electric field $E(\vec{r})$ was numerically calculated on a cross-section of the resonators by the use of a finite-element (FEM) solver. The actual CPW profile was obtained through microscopy on neighboring devices cross sections, see Figure \ref{SEMfig}. The sinusoidal voltage dependence along the length of the resonator is also considered when fitting Eqn. \ref{eqLoss}. 

In the calculations we used the following assumptions: fist, that the conductor-substrate (C-S) interface consists of a 2 nm thick SiN$_x$ layer (confirmed from pre-sputtering ellipsometer measurements\citep{Vissers2010}) with a relative dielectric constant $\epsilon_r = $7.6; second, that the substrate-vacuum (S-V) interface is a 3 nm thick layer with $\epsilon_r =$ 3.9 (dielectric constant of SiO$_2$); third, that the conductor-vacuum (C-V) interface has a $\epsilon_r =$ 10 (value of many metal oxides) and that it is 3 nm thick. We do not know the actual dielectric constant of the C-S interface, but as will be shown later, this is not important as long as $\epsilon_r \gg 1$.    

We find that Eqn. \ref{eqLoss} fits the power dependence of the loss for the F- and Cl-etched resonators if we include a constant loss term in the expression. The origin of this power-independent loss is, as of yet, not determined. One possible reason for the loss could be, despite the double-layer magnetic shielding, trapped magnetic flux in the vicinity of the resonators\cite{Song2009}. We find that it varies by two orders of magnitude between the resonators (ranging from $4\times 10^{-8}$ for resonator Fl2 to $1.8\times 10^{-6}$ for resonator Cl1). However, since this loss is much less than the TLS loss at low powers for a given resonator it can be extracted as a fitting parameter. 

The power dependence of the loss is well fitted by the use of the calculated electric field at both the S-V and the C-S interface by changing the critical electric field. For resonator F1 and F2 we find $E_c$ = 8 to 10 V$\cdot$m$^{-1}$ for the S-V and $E_c = 12$ V$\cdot$m$^{-1}$ for the C-S interface. For resonator Cl1, Cl2 and Cl3 we find $E_c = $25 to 40 V$\cdot$m$^{-1}$ for both interfaces. The fact that the F-etched and Cl-etched resonator loss data do not fit to the same $E_c$ suggests that the loss is dominated by TLSs in different environments. 

The much greater loss observed for the ion-milled resonator IM can, most likely, be attributed to the fence-like structure found on the CPW edges. The fences are formed due to re-deposition of Si onto the edges of the photoresist during the ion-mill process. After the photoresist is stripped off, the fences remains on the edges of the CPW and hence causing the higher loss, see in Figure \ref{SEMfig} (d). 

To analyze the low-power loss, we calculate the filling factors for the S-V, C-V and C-S interfaces. The filling factor, $F_{V}$, of region $V$ is the ratio of the electric energy stored in region $V$ to the total electric energy stored:
\begin{equation}
\label{eqFill}
F_V=\frac{\int_{V}\epsilon_V|E(\vec{r})|^2 \mathrm{d}v}{\int_{V_{tot}}\epsilon(\vec{r})|E(\vec{r})|^2\mathrm{d}v}.
\end{equation} 

The filling factors depend on the geometry of the devices. It has previously been shown that the loss is well explained through filling factor arguments as the resonator trench is changed \cite{Vissers2012}. Assuming that all TLSs are located at the interfaces, the total TLS loss becomes:
\begin{equation}
\label{eqLowLoss}
\delta_{TLS}=F_{\mathrm{S\mbox{-}V}}\delta_{\mathrm{S\mbox{-}V}}+F_{\mathrm{C\mbox{-}V}}\delta_{\mathrm{C\mbox{-}V}}+F_{\mathrm{C\mbox{-}S}}\delta_{\mathrm{C\mbox{-}S}}.
\end{equation}

\begin{figure}[h]
\includegraphics[width=8cm]{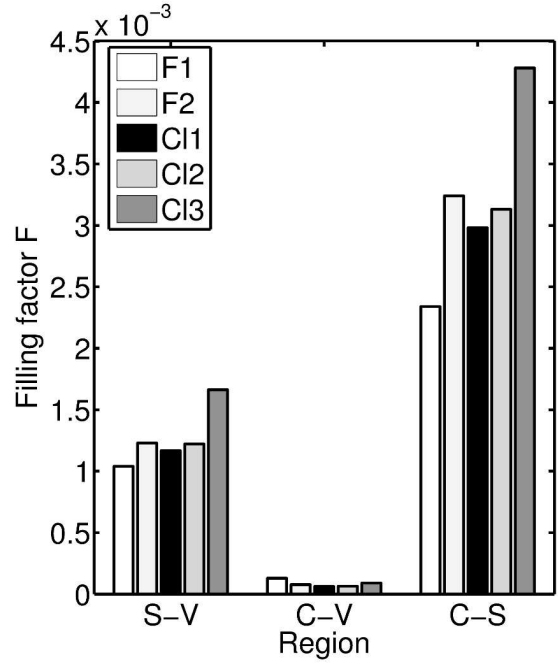}
\caption{\label{fillfig} Calculated filling factors of the substrate-vacuum (S-V), conductor-vacuum (C-V), and conductor-substrate (C-S) interfaces for resonators F1, F2, Cl1, Cl2 and Cl3. Only a minor part of the total electric energy is stored at the interfaces with $\sim$ 90 \% of the total electric energy is stored in the bulk of the Si substrate and only $\sim$ 10 \% is stored in the vacuum. The difference in measured loss between the two Cl-etched resonator Cl1 and Cl3 is well explained by the difference in filling factors of the S-V and C-S interfaces .}
\end{figure}

The filling factors for the different resonators are shown in Figure \ref{fillfig}. We find that the filling factor of the C-V interface is about one order of magnitude smaller than that of the S-V and C-S interfaces. This agrees with the result of Wenner \it et al. \rm\cite{Wenner2011}, indicating that the loss at the conductor surface would have to be one order of magnitude higher than the loss at the other interfaces in order to dominate. This is interesting, since the electric field in the Si substrate and the vacuum are nominally the same, thus, by looking only at Eqn. \ref{eqFill}, one can easily be led to be believe the participation ratios of the top and bottom conductor interfaces should also be nominally the same. The much lower filling factor of the top surface comes from the fact that it is the perpendicular displacement field ($D_{\perp}=\epsilon E_{\perp}$), and not the electric field that has to be continuous at the interface. The imposed boundary condition on $D_{\perp}$ causes the electric field to be either enhanced or suppressed when going to a region with higher $\epsilon_r$ or lower $\epsilon_r$, respectively. Therefore, relatively high $\epsilon_r$ of the conductor interfaces compared to the substrate is desirable to reduce the loss. 

In the calculation of the filling factor of the C-V interface, we assumed $\epsilon_{r}$ = 10. If we instead assume that the top surface is TiO$_x$ ($\epsilon_{r} \gtrsim$ 40), the filling factor would be suppressed even further.   

To compare the TLS losses of the resonators we first subtract the power-independent background loss. We then compare resonators Cl1 and F1, which are fabricated on the same Si wafer. Using Eqn. \ref{eqLowLoss} and assuming that the loss of the C-S interface $\delta_{\mathrm{C\mbox{-}S}}$ is equal for the two resonators, we find that $\delta_{\mathrm{C\mbox{-}S}} \leq 0.4\times 10^{-3}$.  The loss of the F-etched trench is $\delta_{\mathrm{S\mbox{-}V}} \leq 0.9\times 10^{-3}$. Finally, upper and lower bounds for the Cl-etched trench, $1.8\times 10^{-3} \leq \delta_{\mathrm{S\mbox{-}V}} \leq 3.16\times 10^{-3}$ are obtained. If we compare resonators F2 and Cl2 that are co-fabricated on a different wafer, we find that $\delta_{\mathrm{S\mbox{-}V}} \leq 1.8\times10^{-3}$, $\delta_{\mathrm{C\mbox{-}S}} \leq 0.7\times10^{-3}$ and $3.45\times 10^{-3} \leq \delta_{\mathrm{S\mbox{-}V}} \leq 5.3\times 10^{-3}$. 

In both cases the loss of the F-etched trench is lower by at least a factor of two than the loss of the Cl-etched trench. Why the loss varies so much between wafers is not clear. Possible explanations include that the removal of photoresist was done under different conditions (different removers and temperatures) or that the assumptions made for the calculations of the filling factors are not correct. The loss of resonator Cl3 agrees well with the loss of resonator Cl1, considering the filling factors of the S-V and C-S interfaces; see figure \ref{fillfig}. Since the changes in filling factors are almost identical for the two interfaces, we cannot quantify the loss contribution from each region.   

The higher loss of the Cl-etched trench is also accompanied by a very high phase noise of the resonator. We found that the phase noise at 1 kHz of the Cl-etched trench is two orders of magnitude higher than what would been found for F-etched niobium resonators of identical geometry.

There are several possible reasons why the Cl-etched surfaces have a higher loss. One is that the surface layer of the F-etched trench could have a lower dielectric constant, due to deposition of fluorocarbon polymers during the etch. This would decrease the filling factor of the S-V region and hence decrease the contribution to the total loss. Another possible reason for the higher loss is radiation damage. 
We have investigated two methods of decreasing the loss due to the Cl etch. First, we decreased the DC bias during the etch to 0 V to reduce potential ion damage. This did not notability decrease the loss of the resonator.  Secondly we performed a short (11 seconds) F etch. This decreased the measured loss of the Cl-etched resonator by more than a factor of two. These results lead us to believe that the higher loss is most likely due to the result of a lower etch rate and potentially also the presence of boron in the etch gas, which could be implanted into the Si substrate and act as a dopant. The higher etch rate of the F-process is also preferred, because any induced defects get removed at a higher rate, leaving fewer defects at the substrate surface\cite{Fonash1990}.

In conclusion, we have investigated how different etch processes affect the loss of TiN CPW resonators on Si substrates. We found the highest loss for resonators patterned by an Ar-ion mill. We attribute the high loss to a fence-like structure found on the edges of the CPW. The fence structures are formed due to re-deposition of Si onto the photoresist during processing.

The lowest loss was observed for a F-based RIE process. From calculated filling factors we conclude that the loss of the F-processed Si surface is lowered by at least a factor of two than that of the Cl-processed surface. We found that it is the loss originating from the CPW trench that dominates for the  Cl-etched resonators. The trench loss is related to the etch chemistry and not to the DC bias or the amount of trenching.   

These results suggest that even higher quality factors could be achieve by optimizing the etch as well as by post-etch processing of the resonators. However, it is also possible that the remaining loss for the F-etched resonator is dominated by the conductor-substrate interface or even the bulk loss of the substrate. In this case, the loss could be lower by going to larger geometries.           

This work was supported by the NIST Quantum Information
initiative. The views and conclusions contained in this
document are those of the authors and should not be interpreted as representing the official
policies, either expressly or implied, of the U.S. government.

\pagebreak

\begin{table*}[h]
\caption{\label{EtchTab} Parameters for the different etches used in the experiment. An 8" ion gun was used for the ion mill.}
\begin{tabular}{|c|c|c|c|c|c|c|}
\hline
Etch & Pressure [mT] & Gas  & Flow [sccm] & Power [W] & DC bias [V] & Etch rate [nm/s]   \\ \hline

F 	& 100 	& SF$_6$    	& 50 	& 80 	& -68 & TiN/Si: 1/20\\\hline
Cl  	& 30 	& Cl 		& 10   	& 200	& -200    & TiN/Si: 3/3\\
		&		& BCl$_3$	& 30	&		&		  &\\ \hline
Etch & Pressure [mT] & Gas  & Flow [sccm] & Beam current [mA] & Beam voltage [V] & Etch rate [nm/s] \\\hline

IM	& 100 	& Ar    	& 40 	& 40 	& 300 & TiN/Si: 0.033/0.16 \\\hline
\end{tabular}       
\end{table*}

\pagebreak

\begin{table*}[h]
\caption{\label{ResTab} Extracted parameters for the different resonators (see Fig. \ref{lossFig} (a)). The parameters of the different etches are given in Table \ref{EtchTab}. The geometric inductance L$_g$ and kinetic inductance L$_k$ are calculated through the method of Sheen\cite{Sheen1991} \it et al. \rm The best fit is obtained assuming a penetration depth of 245 nm, close to that previously obtained for TiN films\citep{Vissers2010}. The capacitance is obtained from FEM calculations.}
\begin{tabular}{|c|c|c|c|c|c|c|c|c|c|c|c|}
\hline
Resonator & Etch & Depth  & Gap     & Width & Undercut & C & L$_{g}$ & L$_k$ & $\ell$ & f$_r$ & Q$_c$\\ 
          &      &   [nm] &[$\mu$m] &[$\mu$m]& [nm]    &[pF/m]& [$\mu$H/m]  & [$\mu$H/m] & [mm] & [GHz] &\\ \hline
IM & Ar & 650    & 2   & 3      & 0       &    -   & -  &  -  & - & 6.612& 168k\\ \hline
Cl1 & Cl     & 270 & 1.9 & 3.0      & 0       & 176  & 0.42 &  0.52  & 3.318    & 5.58 & 140k \\ \hline
Cl2 & Cl     & 200 & 2 & 3.0      & 0       & 187  & 0.42 &  0.42  & 3.318    & 6.02 & 140k \\ \hline
Cl3 & Cl     & 40    & 2.1  & 2.7   & 0       & 189  & 0.45 &  1.02  & 3.318   & 4.32   & 300k \\ \hline
Fl1 & F & 1200 & 2.3 & 2.4      & 150     & 124  & 0.47  &  0.71 &  3.114  & 6.29 & 1360k \\ \hline
Fl2 & F & 200 & 2 & 3      & 10     & 183  & 0.42  &  0.44 &  3.05  & 6.53 & 602k \\ \hline
\end{tabular}     

\end{table*}

\pagebreak

\pagebreak

\pagebreak

\pagebreak

\bibliographystyle{IEEE}

\end{document}